\documentclass[12pt]{iopart}
\usepackage{graphicx}  
\usepackage{xspace}  

\newcommand{\dphi}{\Delta\phi\xspace}
\newcommand{\pizero}{\mbox{$\pi^{0}$}\xspace}

\newcommand{\dAu}{$d+$Au\xspace}
\newcommand{\pp}{$p$+$p$\xspace}
\newcommand{\rg}{\mbox{$R_g^{Au}$}\xspace}

\newcommand{\rda}{\mbox{$R_{\rm dA}$}\xspace}
\newcommand{\jda}{\mbox{$J_{\rm dA}$}\xspace}

\newcommand{\sqsn}{\mbox{$\sqrt{s_{_{NN}}}$}\xspace}
\newcommand{\pT}{\mbox{$p_T$}\xspace}

\begin{document}

\title[Forward di-hadron measurements in \dAu collisions with PHENIX]{Cold nuclear matter physics at forward rapidities from \dAu collisions in PHENIX}

\author{M Chiu (for the PHENIX Collaboration\footnote{A list of members of the PHENIX
Collaboration can be found at the end of this issue.}) }
\address{Brookhaven National Lab, Upton, NY, 11973 USA}
\ead{chiu@bnl.gov} 

\begin{abstract}
We present measurements by the PHENIX experiment at RHIC of di-hadron
pair production in \dAu collisions where the particles in the pair
are varied across a wide range of pseudorapidity, out to
$\eta = 3.8$.  With di-hadrons, varying the $p_T$ and rapidity of
the particles in the di-hadron pair allows studying any effects as
a function of partonic $x$ in the nucleus.  These di-hadron measurements
might probe down to parton momentum fractions x $\sim$ $10^{-3}$ in the gold nucleus,
where the interesting possibility of observing gluon saturation effects
at RHIC is the greatest.  Our measurements show that the correlated
yield of back-to-back pairs in \dAu collisions is suppressed by up 
to an order of magnitude relative to \pp collisions, and increases
with greater nuclear path thickness and with a selection for lower
x in the Au nucleus.
\end{abstract}


\section{Introduction}

Deuteron-gold collisions at RHIC provide a means to
explore nuclear effects on the initial-state parton
densities in the nucleus, which is vitally important
for understanding the baseline production in heavy-ion
collisions.  RHIC experiments have shown that 
single inclusive hadron yields in the forward (deuteron)
rapidity direction for $\sqsn = 200$ GeV \dAu
collisions are suppressed relative to \pp
collisions~\cite{rda_brahms, rda_star, rcp_phenix}. The mechanism
for the suppression has not been firmly established. 
Many effects have been proposed for this suppression, such as 
gluon saturation~\cite{cgc, monojets}, initial state energy
loss~\cite{vitev2, strikman}, parton recombination~\cite{hwa},
multi-parton interactions~\cite{strikman_parton}, and
leading and higher-twist shadowing~\cite{GSV, vitev}.

One set of measurements that might help to distinguish between the
competing models is forward azimuthally correlated di-hadron correlation
functions, which directly probe di-jet production through their
2$\rightarrow$2 back-to-back peak at $\dphi = \pi$.  This
technique has been used extensively at RHIC and is described in detail
elsewhere~\cite{ida_central_phenix, ida_phenix,jda_paper}.  
The di-hadron results presented here were obtained from \pp and \dAu runs
in 2008 with the PHENIX detector and include a new electromagnetic calorimeter, the Muon Piston
Calorimeter (MPC), with an acceptance of $3.1 < \eta < 3.8$ in pseudorapidity
and $0 < \phi < 2 \pi$.  
Di-hadron measurements can probe more precise ranges of parton $x$ in a gold nucleus
than do single hadron probes (e.g., \rda). At forward rapidities, a single
hadron probe will cover a very broad range of x, $10^{-3}<x_{Au}<0.5$, thus mixing together
the shadowing, anti-shadowing, and even EMC effects~\cite{GSV}.  Azimuthally
correlated di-hadron measurements also enhance the di-jet 
fraction in the event selection, since one selects only the back-to-back hadrons.

\begin{figure}[htbp]
  \begin{center}
  \includegraphics[width=0.55\linewidth]{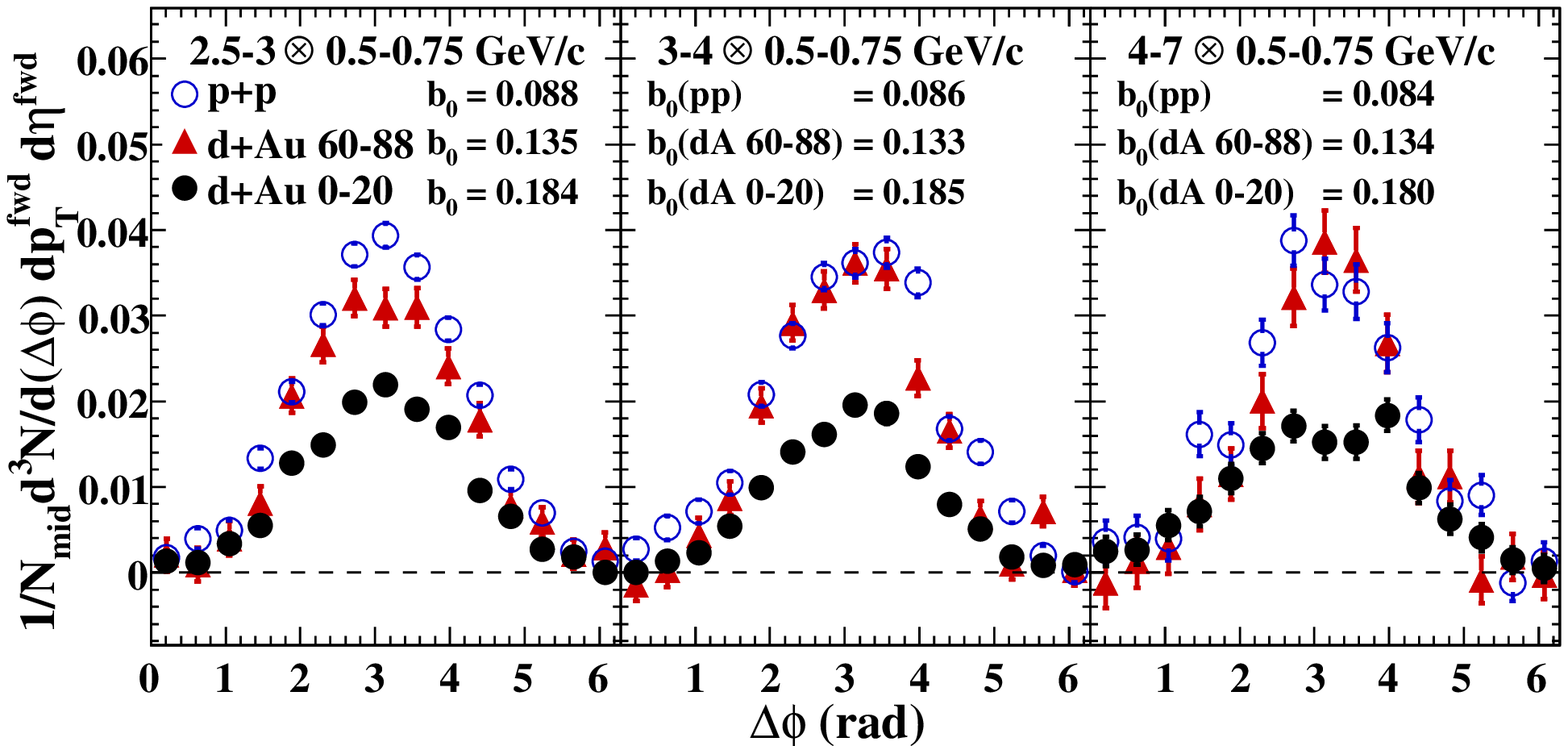} \\
  \includegraphics[width=0.55\linewidth]{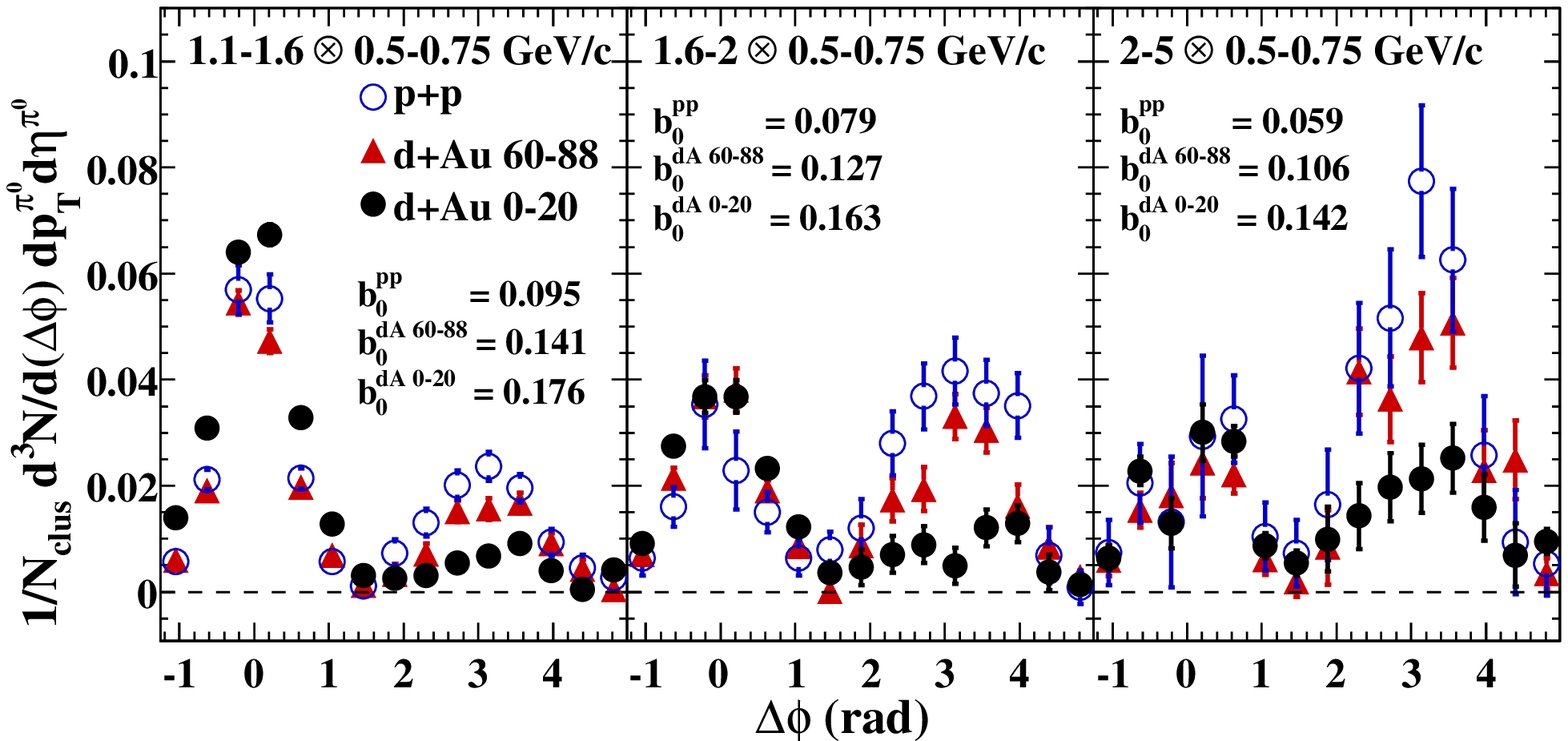}
  \caption{Background-subtracted mid-forward rapidity \pizero-\pizero (top) and 
  forward-forward cluster-\pizero (bottom) per-trigger correlation functions, 
  for \pp (open points), \dAu peripheral  (60-88\%, triangles), 
  and \dAu central collisions (0-20\%, solid points) at $\sqrt{s_{NN}}=200$ GeV.
  The trigger $p_T$ ranges from 1.1 to 7 GeV/c and the associated \pizero's have
  $p_T = 0.5-0.75$ GeV/c.
  Systematic errors of up to 30\% on the near side ($|\dphi|<0.5$) have
  not been shown. The subtracted pedestals, b$_0$, are shown for each case.}  
  \label{fig:mpc_corr}
  \end{center}
\end{figure}

By performing several correlation measurements with
particles at different \pT and rapidities, one can systematically scan different
$x$ ranges with an observable that is enhanced for the leading-order perturbative
QCD component.  Probing the $x$ dependence of the effect is an important test
since most models predict that any effects
should be stronger at smaller $x$. Particles at higher pseudorapidities are produced
from smaller $x$, so measuring hadrons from more forward rapidities
should probe smaller x.

\section{PHENIX MPC \dAu di-Hadron Correlations}

For this analysis, back-to-back \pizero-\pizero or hadron-\pizero pairs are measured
with one particle at mid-rapidity, and the other at forward rapidity.  Back-to-back
cluster-\pizero pairs are also measured where both are in the forward rapidity region.
The clusters are reconstructed from the energy deposit of photons in the MPC,
and are estimated to be at least 80\% dominated by \pizero's, with the remainder coming
from single photons from decays of $\eta$'s and from direct photons.  Further details of the
analysis are available in~\cite{jda_paper}.

As shown in figure~\ref{fig:mpc_corr}, the away-side peak for \dAu central collisions appears significantly
suppressed compared to \pp collisions and peripheral \dAu collisions.
This effect is large for the mid-forward di-hadron correlations
and becomes even larger for the forward-forward correlations.
Within large errors, the Gaussian widths of the away-side correlation peak for the
mid-forward di-hadron correlations remain the same
between \pp and central \dAu.  For the forward-forward case, uncertainties in the pedestal level
from the underlying event and the strong suppression of the away-side peak make extracting the width
unreliable.  For this case, the away side peak width in central \dAu collisions is allowed to vary up to twice
as much as in \pp when accounting for this systematic uncertainty.

\begin{figure}[htbp]
  \begin{center}
  \includegraphics[width=0.6\linewidth]{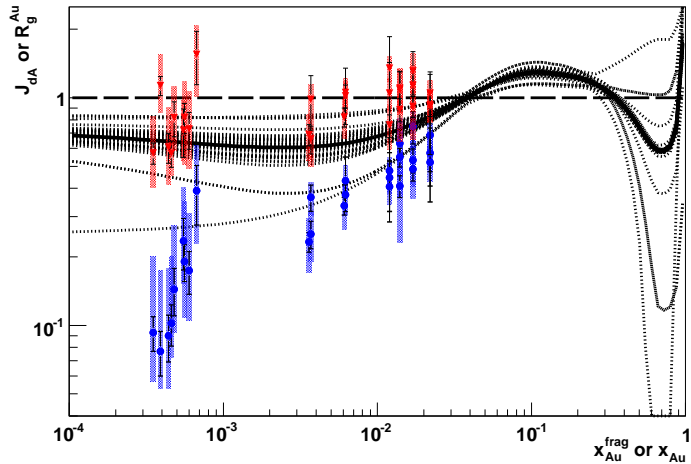}
  \caption{$J_{dA}$ versus $x_{Au}^{frag}$ for peripheral (60-88\%, in red) and central (0-20\%, in blue) \dAu collisions at
  $\sqrt{s_{NN}}=200$ GeV, and compared to the EPS09 LO $R_g^{Au}(x_{Au})$ curves at a scale $Q^2$ = 4 GeV$^2$~\cite{eps09}. }
  \label{fig:jdax}
  \end{center}
\end{figure}

The observed suppression is quantified by studying the
relative yield, $J_{dA}$~\cite{jda_orig}, of correlated back-to-back hadron pairs in $d+$Au 
collision compared to \pp collisions scaled with the average
number of binary nucleon collisions $\langle N_{coll} \rangle$,
\begin{equation}
\jda = \frac{1}{\langle N_{coll} \rangle}\frac{\sigma_{dA}^{pair}/\sigma_{dA}}{\sigma_{pp}^{pair}/\sigma_{pp}}.
\end{equation}
This is simply the analog of the usual nuclear modification factor \rda but for hadron pairs.
The $\sigma$ are the \pp or \dAu inelastic cross-sections, while $\sigma^{pair}$ is the cross-section for 
di-hadron pair production.
The \jda is calculated using the correlated away side peak after subtracting the pedestal $b_0$.
\jda decreases with increasing number of binary collisions,
$\langle N_{coll} \rangle$, or equivalently with increasing nuclear
thickness.  The suppression also increases with decreasing particle
$p_T$ and is significantly larger for forward-forward hadron pairs
than for mid-forward pairs. The
observed suppression of $J_{dA}$ versus nuclear thickness, $p_T$ and
$\eta$ points to large cold nuclear matter effects arising at low
parton momentum fractions x in the nucleus probed by the deuteron. This trend is seen
more clearly in Fig.~\ref{fig:jdax} where
$J_{dA}$ is plotted versus $x_{Au}^{frag}=
(\langle p_{T1}\rangle e^{-\langle \eta_1\rangle} + 
\langle p_{T2}\rangle e^{-\langle \eta_2\rangle})/\sqrt{s_{NN}}$
 for all pair selections in $\eta$ and $p_T$.
In the case of 2$\rightarrow$2 leading order (LO) processes,  
the variable  $x_{Au}^{frag}$ is lower than $x_{Au}$ by the mean 
fragmentation fraction, $\langle z \rangle$, of the struck parton 
in the Au nucleus. Since $x_{Au}^{frag}$ is an entirely experimental
defined quantity, it should be reproducable in any theoretical framework.

\section{Discussion}

In a leading order pQCD picture, the variable \jda is
\begin{equation}
J_{dA} = \frac{\sigma_{dA}^{pair}/\sigma_{dA}}{\langle N_{coll} \rangle\, \sigma_{pp}^{pair}/\sigma_{pp}} \approx
\frac{f^a_d(x^a_d)\otimes f^b_{Au}(x^b_{Au})\otimes\hat{\sigma}^{ab\rightarrow cd}\otimes\mathcal{D}(z_c,z_d)}
{\langle N_{coll} \rangle \, f^a_p(x^a_p)\otimes f^b_p(x^b_p)\otimes\hat{\sigma}^{ab \rightarrow cd}\otimes\mathcal{D}(z_c,z_d)}
\end{equation}
for partons a+b going to outgoing jets c+d, which then fragment to hadrons with
longitudinal fractions $z_c$, $z_d$.  
In the above convolutions over p+p and d+A, most of the terms are expected to be roughly similar
between p+p and d+Au except for the nuclear gluon pdf.  
Naively, \jda might be largely dominated by the modification to the nuclear gluon parton
distribution function (pdf's), since most of the events with di-hadrons at forward rapidities consist of
a high-x parton from the deuteron and a low-x gluon from the gold nucleus.
Assuming this to be true then $J_{dA} \sim \rg = G_{Au}(x,Q^2)/A\,G_p(x,Q^2)$
In figure~\ref{fig:jdax} the \jda values are overlaid with the EPS09 \rg
curves~\cite{eps09}.  The \jda values for the peripheral bins
are above the best fit EPS09, while the central bins are below.  
The EPS09 curves are taken largely from nuclear deep
inelastic scattering and represent an averaged value of \rg over all centralities.
The \jda values for the most central bin at the lowest x are well below
the EPS09 curves.  This is qualititatively consistent with
the expectations for the Color-Glass Condensate~\cite{cgc}, which
posits an extreme form of shadowing due to the onset of gluon
saturation.

If nature is kind and this data can be interpreted in terms
of a simple LO pQCD picture, it may be possible to extract \rg,
which is extremely important for understanding the quark gluon
plasma since it forms the baseline
for production in heavy ion collisions.  In addition, if the large
suppression of \jda  observed in central \dAu collisions is from
gluon saturation, it may be possible to study the dependence of that
saturation on the thickness of the nucleus.  One possible test of whether
these ideas are correct would be to use extractions of \rg from
this data to predict $J/\Psi$ data in \dAu collisions from PHENIX.


\end{document}